# Hybrid metal-dielectric nanocavity for enhanced light-matter interactions

**Yousif A. Kelaita**[1,*], **Kevin A. Fischer**[1], **Thomas M. Babinec**[1], **Konstantinos G. Lagoudakis**[1], **Tomas Sarmiento**[1], **Armand Rundquist**[1], **Arka Majumdar**[2], and **Jelena Vučković**[1]

[1]*Department of Electrical Engineering and Ginzton Laboratory, 348 Via Pueblo, Stanford University, Stanford, CA 94305, USA.*
[2]*Department of Electrical Engineering and Department of Physics, 185 Stevens Way, University of Washington, Seattle, WA 98195, USA.*
*\*ykelaita@stanford.edu*

**Abstract:** Despite tremendous advances in the fundamentals and applications of cavity quantum electrodynamics (CQED), investigations in this field have primarily been limited to optical cavities composed of purely dielectric materials. Here, we demonstrate a hybrid metal-dielectric nanocavity design and realize it in the InAs/GaAs quantum photonics platform utilizing angled rotational metal evaporation. Key features of our nanometallic light-matter interface include: (i) order of magnitude reduction in mode volume compared to that of leading photonic crystal CQED systems; (ii) surface-emitting nanoscale cylindrical geometry and therefore good collection efficiency; and finally (iii) strong and broadband spontaneous emission rate enhancement (Purcell factor ~ 8) of single photons. This light-matter interface may play an important role in quantum technologies.

## 1. Introduction

A prototypical cavity quantum electrodynamics (CQED) system that consists of a quantum emitter (QE) in an optical cavity is described by three rates, the emitter decay rate γ, the cavity field decay rate κ, and the coherent emitter-cavity coupling rate $g$ [1]. The coherent coupling rate scales as $g \propto 1/\sqrt{V}$, where $V$ is the cavity mode volume, and therefore increasing field localization has long been recognized as a path to boosting emitter-field coupling strength [2-5]. For example, whispering gallery GaAs microdisk resonators exhibit mode volumes $V \sim 5(\lambda/n)^3$ and coupling to InAs quantum dots with rate $g/(2\pi) \sim 2-3\ GHz$ [6]. Planar photonic crystal nanocavities possess an order of magnitude smaller mode volume $V \sim 0.5(\lambda/n)^3$ with light-matter coupling strengths up to $g/(2\pi) \sim 40\ GHz$ [7-9].

Moving beyond this nanophotonic CQED regime and achieving $g/(2\pi) > 40\ GHz$ requires coupling a single QE to a nanoscale cavity characterized by deep sub-wavelength optical confinement [10,11]. While recent work theoretically demonstrates the ability to reach ultra-small V with dielectric cavities [12], experimental efforts have focused primarily on nanometallic cavities and, in particular, plasmonic cavities that can shrink the effective optical wavelength. For example, continued work on the now common nanoparticle-on-mirror (NPOM) geometry [13] has led to room-temperature strong coupling between the NPOM cavity and organic fluorophores [14]. However, there have been no related reports involving solid-state emitters such as defect centers in crystal lattices or self-assembled emitters such as the InAs quantum dot (QD), which is of great interest for quantum light sources thanks to its large internal quantum efficiency and short radiative lifetime [15]. In order to reach this regime with such emitters, we require a cavity that can achieve ultra-small V without resorting to plasmonic effects, as solid-state emitters have been shown to quench in their presence [16].

In this article, we investigate a hybrid metal-dielectric CQED system that is a variation of recent theoretical proposals [3,17] and an integrated solid-state alternative to early demonstrations with other emitters such as dye molecules and colloidal quantum dots [18-22]. Notably, our modification allows for surface emission and hence good collection efficiency compared both to the bulk and to fully-embedded geometries that require optical addressing through a substrate. We propose this geometry as a general platform for quantum photonics and discuss the fabrication challenges inherent to surface-emitting metal-dielectric material interfacing. Finally, we demonstrate broadband enhancement of the rate of spontaneous emission of a QD as well as single-photon emission as a proof-of-concept for the platform's potential for strong light-matter interactions. The proposed platform may play an important role in the studies of fundamentals of cavity quantum electrodynamics [23] as well as in its application to optical interconnects [24] and quantum information processing [25-27].

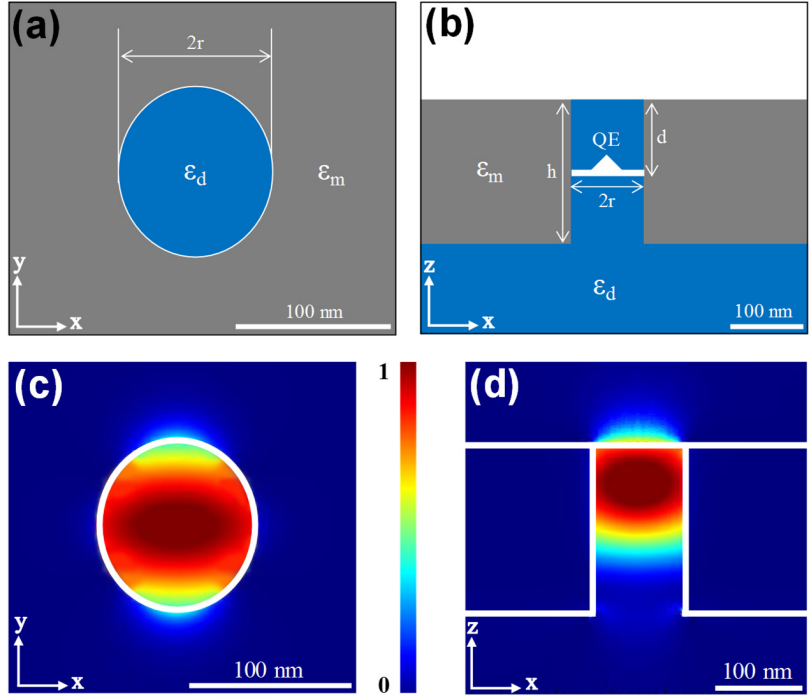

Fig. 1. Schematic depiction of the hybrid metal($\varepsilon_m$)-dielectric($\varepsilon_d$) nanocavity from (a) top-down and (b) side views. FDTD simulation of the nanocylinder cavity field intensity ($|E_x|^2$) for the x-polarized $TE_{11}$ waveguide mode viewed in the nanocylinder cross-section (c) and along the nanocylinder axis (d).

## 2. Structure and modeling of the hybrid metal-dielectric nanocavity

Our nanocavity consists of a dielectric nanopillar surrounded by a metallic film on all sides but left bare on the top surface as shown in Figs. 1(a) and 1(b). This nanopillar behaves as a cylindrical waveguide with guided TE and TM modes that are confined vertically by an impedance mismatch between the air, waveguide, and underlying substrate. This mismatch gives rise to a weak Fabry-Perot effect that leads to a standing wave in the vertical direction. Using the finite-difference time-domain (FDTD) method, we simulate the fundamental $TE_{11}$ nanocylinder mode of a GaAs (n = 3.46) nanopillar with r = 50 nm and h = 200 nm that is surrounded by a Ag [28] film, as shown in Figs. 1(c) and 1(d). We note that the in-plane field profile mimics that of the textbook cylindrical waveguide, while the vertical profile demonstrates confinement in the top half of the pillar, as expected given the smaller impedance mismatch between the effective index of the waveguide and air. The pillar supports degenerate $\hat{x}$ and $\hat{y}$ polarized $TE_{11}$ waveguide modes but we concentrate our ensuing discussion on one polarization. Finite-difference time-domain simulations show that the nanocavities possess resonances with optimal quality factor Q ~ 25 and ultra-small mode volumes $V \sim 0.025 \, (\lambda/n)^3$ at the emission wavelength according to $V = \iiint \frac{\partial(\omega\varepsilon(r))}{\partial\omega} |\vec{E}(\vec{r})|^2 \, dV \, / \, \max\left[ \frac{\partial(\omega\varepsilon(r))}{\partial\omega} |\vec{E}(\vec{r})|^2 \right]$ [29].

A QE embedded within the nanocavity may decay via emission of a photon polarized into either of the orthogonal $TE_{11}$ nanocylinder modes. Moreover, the spontaneous emission rate is enhanced from the bulk rate due to the enhanced density of optical states (Purcell effect). The cavity exhibits a maximum achievable Purcell factor $F_p = \tau_{Bulk} / \tau_{Cavity} \sim 35$ according to the expression $F_p = \frac{3}{4\pi^2} \left(\frac{\lambda_c}{n}\right)^3 \left(\frac{Q}{V}\right)$, which agrees with the FDTD simulation in Fig. 2. In this case, application of the dipole approximation is valid despite the small size of the cavity since the gradient of the field at the QD position is negligible when near the field maximum [30].

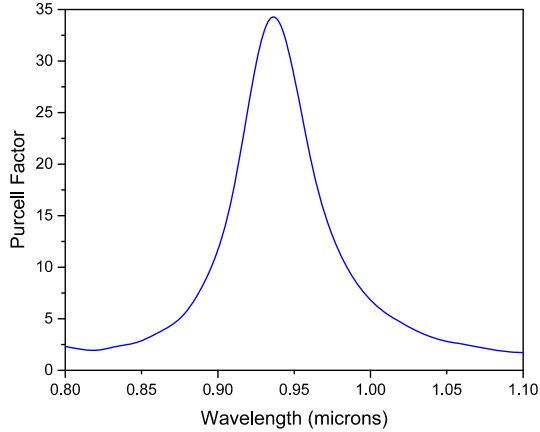

Fig. 2. FDTD simulation of Purcell factor as a function of emission wavelength for an r = 50 nm, h = 200 nm cavity, demonstrating an approximate Q ~ 25.

In practice, solid-state QEs can spectrally cover a wide range of both the visible and near-infrared regimes of the electromagnetic spectrum. In addition, emitters can be spatially located at arbitrary depths in their host substrates, but only in the case of growth processes allowing for precise control of depth such as molecular beam epitaxy. For strong light-matter interaction, the nanocavity must be designed such that the fundamental mode is both spectrally aligned with the emitter's optical transition and spatially concentrated such that the field maximum lies at the emitter's location. In order to satisfy these conditions for an arbitrary emitter, we present detailed FDTD simulations showing the mode-dependence on key geometric features: nanopillar radius and height.

First, we investigate tuning of the resonance as a function of the nanopillar radius. The resonance dramatically red-shifts for increasing radius, consistent with an increasing waveguide effective index, as shown in Fig. 3(a). Hence, small radius modifications can be utilized as a coarse adjustment for spectral alignment. The radius can be arbitrarily small but will be practically limited by fabrication constraints and quenching of emitters if they are positioned too close to interfaces. On the other hand, we note that beyond a particular radius that depends on the specific material system, the higher-order cylindrical waveguide modes will be above cutoff and hence the nanocavity will enter the undesired multi-mode regime. Operation in the higher order mode regime results in an enlarged mode volume and has the potential to shift the spatial field maximum off of the QE target depth.

We then perform a similar analysis by varying pillar height in Fig. 3(b). Varying pillar height can be utilized for emitter spatial alignment in the vertical direction given that the field is confined to the top half of the pillar. Hence, a taller pillar may be used for emitters that are deeper into the substrate and vice-versa. Spectrally, we observe a redshift for increasing height, which is consistent with increasing the length of a Fabry-Perot cavity in line with our intuitive model. This redshift is less pronounced than for an equivalent change in pillar diameter, and therefore can be considered a finer adjustment for spectral alignment. Realistically, the pillar height can only be as small as the diameter of the pillar, given that the Fabry-Perot terminated waveguide picture breaks down below a 1:1 aspect ratio and the nanocavity will no longer support cylindrical waveguide modes. On the other hand, the high aspect ratios required for increased pillar heights present increased demands on multiple steps in the fabrication process including the etch and metallization.

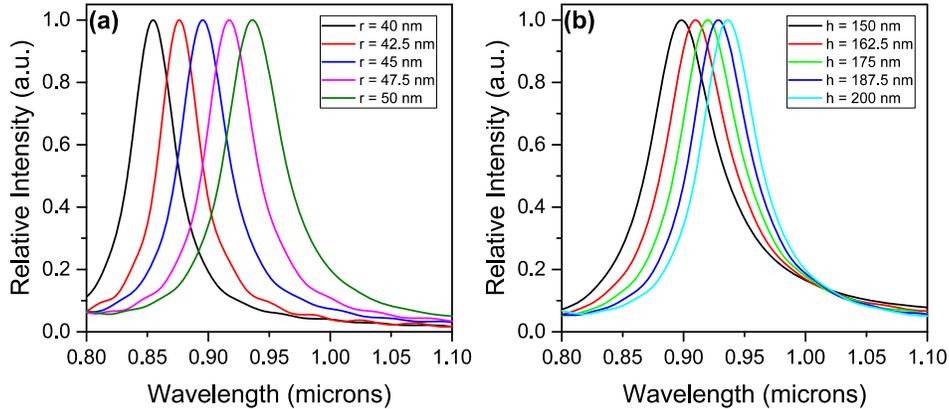

Fig. 3. FDTD parameter study of (a) pillar resonances as a function of radius r, for pillar height h = 200 nm and (b) as a function of pillar height h, for a fixed pillar radius r = 50 nm. These results demonstrate cavity resonance shift as a function of the pillar parameters in line with the intuitive description of the nanocavity as a Fabry-Perot terminated cylindrical waveguide. Here, refractive index of the pillar is n = 3.46 corresponding to GaAs at a temperature of 10 K in the studied wavelength range, and the pillar coating is Ag with refractive index from optical constant data [28].

Therefore, there exists a tradeoff between maintaining the desired electromagnetic landscape and allowing for a design that can be fabricated reliably. Despite this tradeoff, there remains a large parameter space of radii and heights that can be tailored to an emitter of interest as evidenced above.

## 3. Results and Discussion

### 3.1 Nanofabrication strategies and challenges

The proof-of-concept structure is realized in the well-established InAs/GaAs quantum dot platform. Our QE is a bound electron-hole pair (neutral exciton $X^0$) confined in a self-assembled InAs quantum dot grown by the Stranski-Krastanov process in a molecular beam epitaxy system. Individual InAs quantum dots are roughly ~ 20 nm in lateral cross section and about 3 nm tall, and are located ~ 100 nm below the top surface of the GaAs wafer. We note that this process can be adapted to any material system that exhibits QEs provided a reliable anisotropic etch recipe exists.

The metal-dielectric nanocavity is defined in the GaAs wafer using a 100 kV electron-beam lithography system (JEOL 6300-FS) and a process based on a negative-tone electron-beam resist (ma-N 2405). The high acceleration voltage allows for resolution of circular pillars with diameters as low as 70 nm due to minimal forward scattering in the resist layer. The negative-tone process allows for fast and compact exposure of devices on a chip, minimizing deleterious backwards scattering that can lead to overexposure. The devices are exposed in a periodic pattern and align to randomly positioned QDs in a purely stochastic manner.

The resist pattern is transferred to the device layer with an inductively-coupled plasma reactive-ion etch process (Oxford Instruments PlasmaPro). The devices are etched using $BCl_3$ at a flow rate of 2 sccm and Ar at a flow rate of 28 sccm, with a chamber pressure of 2 mTorr and driving powers of 60 W of RF power and 450 W of ICP power. This balance of physical and chemical etching results in a highly vertical sidewall etch profile as shown in Fig. 4(a).

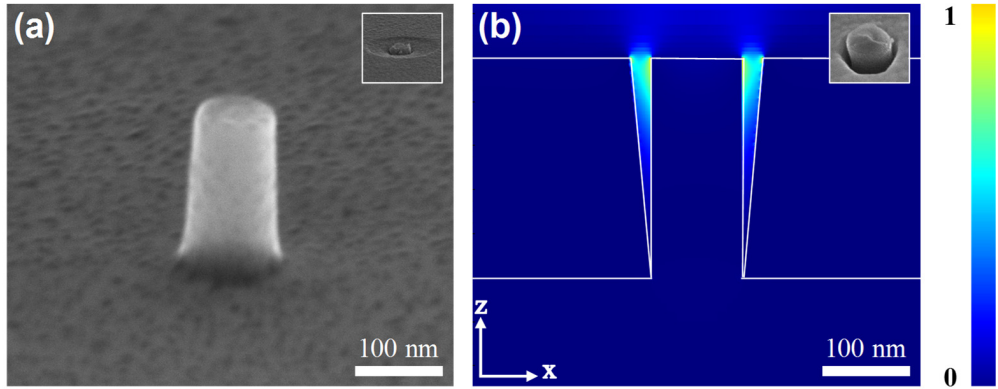

Fig. 4. (a) Colorized SEM micrograph of a bare pillar with resist mask removed to show the vertical etch profile. For actual devices, the resist mask is used to lift off metal after the angled rotational deposition process, leaving an uncapped pillar that is surrounded by metal (inset of a). (b) $|E_x|^2$ of a pillar with a 20 nm air gap to the metal wall, showing field localization in the gap, with an SEM micrograph of a fabricated pillar with standard top-down metal deposition that exhibits the air gap (inset of b).

The nanocavity requires a surrounding Ag film that conformally coats the sidewalls of the pillar. The presence of any air gap between metal and dielectric precludes the $TE_{11}$ mode and leads to field localization in air and hence diminished interaction with embedded solid-state QEs at their frequencies of emission, as shown in Fig. 4(b). Conventional electron-beam or thermal metal evaporation cannot satisfy this requirement as it is anisotropic and hence experiences a self-shadowing effect as metal accumulates on the resist mask used for liftoff, as shown in the inset of Fig. 4(b). For this reason, a common strategy in other metal-dielectric devices is to remove the resist mask before capping the entire device in a metal film. However, this correlates with drastically reduced extraction efficiencies due to collection through the substrate. Here we devise and demonstrate an approach to conformally coat the sidewalls of a pillar without coating the top to allow for surface emission. We evaporate Ag using a custom electron-beam metal evaporation system that uniquely allows for rotation about an axis that can tilt along with the substrate. We evaporate onto etched devices that are rotating about an axis tilted at 45°, allowing for metal to accumulate directly on the walls of the pillar as shown in Fig. 4(a). Finally, the remaining resist mask can easily be removed using n-methyl-2-pyrrolidone in an ultrasonic bath.

We note that for any device with a vertical sidewall and a high aspect ratio, metal coating with evaporation or sputtering will likely result in voids due to the aforementioned shadowing phenomenon even for fully embedded structures. We propose that our scheme for angled rotational deposition be used for any device requiring a conformal metal coating.

### 3.2 Optical characterization and photon statistics measurements

For the following measurements, the InAs quantum dot is contained within a GaAs nanopillar of radius r ~ 40 – 50 nm and height h ~ 200 nm that is surrounded by a Ag film. The QD is located at a local field intensity of $E/E_{max} = 0.8$ due to suboptimal vertical positioning. The measurements are taken in an open-flow helium cryostat at a temperature of 10 K, though we note that this platform can easily be extended to room-temperature operation given a QE that maintains electron confinement at high temperatures.

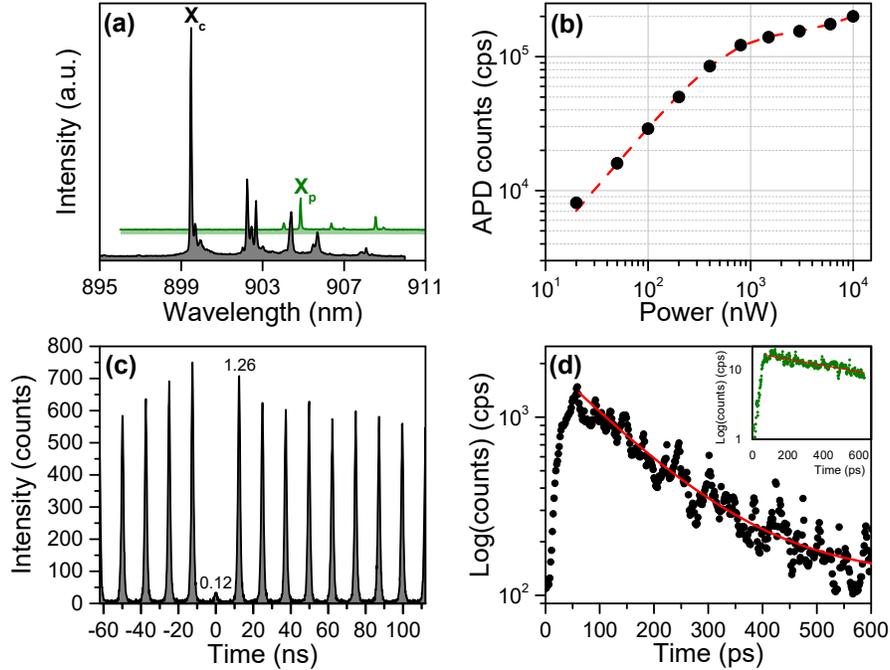

Fig. 5. (a) Representative photoluminescence spectrum of a nanocavity containing a single quantum dot. The single exciton (X) line is highlighted for coupling to the nanometallic cavity $X_c$ (black) and a reference bare pillar $X_p$ (green) that are offset horizontally (0.8 nm) and vertically (arbitrary) for clarity. (b) Total number of single photons collected per second (black circles) for quasi-resonant excitation of the exciton line $X_c$. The horizontal axis corresponds to incident power, measured before the objective lens, and the source is excited with pulses at an 80MHz repetition rate. The data is fit to a saturation model (dashed red line). (c) Intensity autocorrelation $g^{(2)}(\tau)$ measurements filtered on the single exciton line well above saturation powers show strong photon antibunching. (d) Fluorescence decay measurements (black circles) with an exponential decay fit (red line) taken with a streak camera for an exciton line in a hybrid metal-dielectric nanocavity, showing 8-fold enhancement in the radiative decay rate (spontaneous emission) over an exciton line in a bare pillar (inset).

Roughly one in ten fabricated cavities exhibits multiple emission lines from a variety of QD states, including the neutral exciton ($X^0$), charged exciton ($X^-$, $X^+$) and bi-exciton (XX). The intensity of these quantum dots for well-coupled QD-cavity systems is quite bright with approximately ~ 8 – 10 X improvement compared to individual lines addressed in a bare nanopillar, as shown in Fig. 5(a). Due to the inability to isolate single QD exciton lines in bulk at the growth densities used in this experiment, we perform the reference measurement on a bare pillar of radius ~ 200 nm such that a single exciton line can be isolated without affecting the photonic density of states. In some of the nanometallic cavity devices (not used for the presented experiments), significant broadening of all transitions suggests that the quantum dots were located in close proximity to the GaAs/Ag interface.

With the numerical aperture NA ~ 0.75 lens used in our experimental setup, FDTD simulations indicate that the lens collects $\eta_{col,c}$ ~ 7% of the photons coupled to the nanocavity mode assuming negligible loss due to metal absorption of photons emitted by the cavity mode. For a bare nanopillar with the size used in the reference measurements, the same lens collects $\eta_{col,p}$ ~ 3% of photons. Beyond this improvement in collection efficiency, any further enhancement provided by the nanometallic cavity relative to a bare pillar is subsequently evaluated as a Purcell effect. We note that FDTD simulations show both the cavity and a bare pillar exhibit improvements over the collection efficiency $\eta_{col,b}$ ~ 1.5% of a QD dipole in unprocessed bulk material.

After demonstrating that we may optically isolate a single exciton line in the cavity, we now characterize its properties as a single photon source. To do so, we measure the number of single photon counts detected per second (CPS) as a function of incident optical pump power (P, measured before the objective lens) under quasi-resonant, pulsed excitation and present this data in Fig. 5(b). We perform pulsed excitation with an 80MHz repetition rate. As is expected for a two-level system, we observe that the number of single photon counts increases linearly at low pump power and saturates at high powers [31]. Fitting photon counts data to a saturation model with the form $CPS_{SAT} = CPS_{BG} + CPS_{LIN} \cdot P + CPS_0 (1 - e^{-P/P_{SAT}})$ gives $(CPS_{BG}, CPS_{LIN}, CPS_0, P_{SAT}) = (860, 6.4\, nW^{-1}, 1.35 \times 10^5, 435\, nW)$. The constant background comes from detector dark counts, while linear background is due primarily to leakage from the pump laser. Considering that we utilize pulsed excitation with repetition period much longer than the excitonic lifetime, the saturation power is not dependent on the modified excitonic lifetime. However, the single photon counts in continuous-wave saturation are thanks to strong Purcell enhancement in addition to an improvement in collection efficiency coming from the structure geometry. We note that with a detector efficiency of ~ 40%, a collection efficiency of 7%, a setup transmissivity of ~ 22%, and ~ 105kCPS collected at saturation from the orthogonal y-polarized cavity mode (data not shown), the total count rate of 240kCPS remains under the theoretical count rate of 500kCPS for an 80MHz repetition rate laser. We attribute the remaining discrepancy to metal absorption of photons emitted from the cavity mode on the order of ~ 50%, leading to an overall source efficiency $\eta$ ~ 4%. With superior metal deposition techniques that lead to higher quality films, the source efficiency can approach the theoretical efficiency of 7%.

In Fig. 5(c), measurement of the intensity autocorrelation function $g^{(2)}(\tau) = \langle I(t)T(t+\tau) \rangle / \langle I(t) \rangle^2$ filtered on the neutral exciton line $X_c$ for quasi-resonant excitation in a Hanbury-Brown and Twiss (HBT) experiment shows strong anti-bunching $g^{(2)}_{X_c} \sim 0.12$, further confirming the presence of a single QE coupled to the cavity mode. To extract this number as well as $g^2(0)$ for the nearest peak, we compare the counts at each peak to the counts for a peak far from the zero-delay point, as nearby peaks experience bunching due to QD-blinking [32]. While $g^2(0)$ is quite low, the value is non-zero due to re-excitation under quasi-resonant excitation of the neutral exciton from other states in the QD complex, such as the biexciton and charged exciton [33,34].

We conclude with a characterization of the dynamics of single photon generation by the nanocylinder cavity. First, we establish the roughly unmodified spontaneous emission decay $\tau_{Bulk} = 1.02 \pm 0.033$ ns via pulsed excitation of an exciton in a large non-metallized nanopillar device and imaging on a streak camera in the inset of Fig. 5(d). Next, we performed the same measurement for the exciton coupled to the mode of our nanometallic cavity. We observe a modified QD lifetime of $\tau_c = 142 \pm 7$ ps in Fig. 5(d). Justified by the large single photon counts observed from this device, we attribute the full intensity decay rate modification to the radiative effects, and estimate Purcell factor $F_{p,c} \sim 7 - 8$ for a single quantum dot. Because the metallic losses only affect the damping of the cavity mode (and not of the emitter itself, based on the unmodified emitter linewidth in a nanometallic cavity), the calculation of the Purcell factor can be performed in the same way as for a lossy dielectric cavity. Such a Purcell factor would already redirect nearly all of the QD spontaneous emission into the nanocavity mode, with the spontaneous emission coupling factor $\beta = \dfrac{F_p}{F_p + 1}$ being near unity.

## 4. Conclusion

In conclusion, we have demonstrated a novel and versatile light-matter interface for single QEs featuring a hybrid metal-dielectric nanocavity. Key features of our platform include very small mode volume, a surface emitting nanoscale cylindrical geometry, and strong and broadband spontaneous emission rate modification via metallic confinement. Furthermore, we demonstrated a method for conformal metal deposition even for structures that require a bare top for surface emission that has broad applicability to various designs. In the future, even smaller mode volumes can be achieved using coaxial structures that would equally benefit from the fabrication improvements demonstrated here. Finally, this light-matter interface can also be implemented in emerging room-temperature quantum systems such as diamond [35] and silicon carbide [36]. With sufficient QE density in such systems, this nanocavity can be a candidate for room-temperature strong coupling between an ultra-small mode volume cavity and a solid-state QE.

## Funding

U.S. Air Force of Scientific Research Multidisciplinary University Research Initiative (AFOSR-MURI) on Quantum Metaphotonics and Metamaterials (FA-9550-12-1-0488).

## Acknowledgements

Y.A.K. acknowledges support from the Department of Defense (DoD) through the National Defense Science & Engineering Graduate Fellowship (NDSEG) Program and from the Art and Mary Fong Stanford Graduate Fellowship (SGF). K.A.F. acknowledge support through the NDSEG and Lu SGF. T.M.B. acknowledges support through the Nanoscale and Quantum Science and Engineering Postdoctoral Fellowship at Stanford. We are grateful to Prof. Jim Harris (Stanford University) for use of his MBE facilities. Also, we acknowledge Rich Tiberio, Cliff Knollenberg, and Tom Carver with the Stanford Nano Shared Facilities (SNSF) for assistance in fabricating samples.